\documentclass[bjps]{imsart}

\RequirePackage{amsthm,amsmath,amsfonts,amssymb}
\RequirePackage[authoryear]{natbib}
\RequirePackage[colorlinks,citecolor=blue,urlcolor=blue]{hyperref}
\RequirePackage{graphicx}

\startlocaldefs
\theoremstyle{plain}

\theoremstyle{remark}

\endlocaldefs

\begin{document}

\begin{frontmatter}
\title{An expectation-maximization algorithm for estimating the parameters of the correlated binomial distribution}
%\title{A sample article title with some additional note\thanksref{t1}}
\runtitle{An EM for correlated binomial distribution}
%\thankstext{T1}{A sample additional note to the title.}

\begin{aug}
\author[A]{\inits{A.}\fnms{Andrea} \snm{Bennett, BS}\ead[label=e1]{first@somewhere.com}},
\and
\author[A]{\inits{M.}\fnms{Min} \snm{Wang, PhD}\ead[label=e2,mark]{min.wang3@utsa.edu}}
%%%%%%%%%%%%%%%%%%%%%%%%%%%%%%%%%%%%%%%%%%%%%%
%% Addresses                                %%
%%%%%%%%%%%%%%%%%%%%%%%%%%%%%%%%%%%%%%%%%%%%%%
\address[A]{Department of Management Science and Statistics, The University of Texas at San Antonio, San Antonio, Texas, USA
\printead{e2}}

\end{aug}

\begin{abstract}
The correlated binomial (CB) distribution was proposed by \citeauthor{Luce:1995} (Computational Statistics \& Data Analysis 20, 1995, 511–520) as an alternative to the binomial distribution for the analysis of the data in the presence of correlations among events. Due to the complexity of the mixture likelihood of the model, it may be impossible to derive analytical expressions of the maximum likelihood estimators (MLEs) of the unknown parameters. To overcome this difficulty, we develop an expectation-maximization algorithm for computing the MLEs of the CB parameters. Numerical results from simulation studies and a real-data application showed that the proposed method is very effective by consistently reaching a global maximum. Finally, our results should be of interest to senior undergraduate or first-year graduate students and their lecturers with an emphasis on the interested applications of the EM algorithm for finding the MLEs of the parameters in discrete mixture models.
\end{abstract}


\end{frontmatter}

\section{Introduction}

In an introductory probability theory and statistics course, the binomial distribution appears naturally as one of the most important discrete probability distributions. It stands for the number of successes in a sequence of $n$ independent experiments when carrying out a series of Bernoulli events. The probability mass function (PMF) of the Binomial random variable Y with Bernoulli probability $p \in [0,1]$ is given by
\begin{equation}\label{binomial}
P(Y = y \mid n, p) = {{n}\choose{y}} p^y (1 - p)^{n-y},
\end{equation}
for $y = 0, 1, \cdots, n$. It is well known that the binomial distribution relies on a key assumption that each event is independent of every other event, whereas such an assumption may not be satisfied in some practical applications. For example, as in a commercial soybean selection study by \cite{Dini:Tuti:2010}, there exists a correlation between any two plants in each pot when competing about the quantity of nutrients. Suppose that we are interested in calculating the probability of choosing a good plant from a pot, and the binomial distribution may not be appropriate since the assumption of independence in the pot may not be valid. This observation motivated some researchers to generalize the binomial distribution in various ways; see for example, \cite{Rudo:1990}, \cite{Mads:1993}, \cite{Luce:1995}, \cite{Pire:Dini:2012}, to name just a few.

It deserves mentioning that \cite{Luce:1995} proposed the correlated binomial (CB) distribution, denoted by CB$(n, p, \rho)$, which is derived based on a combination of the binomial distribution and a modified Bernoulli distribution. Specifically, the PMF of the CB distribution with the parameters $p$ and $\rho$ is given by
\begin{equation}\label{cb:01}
P(Y = y \mid n, p, \rho) = {{n}\choose{y}} p^y (1 - p)^{n-y}(1 - \rho)I_{A_1}(y) + p^{y/n} (1-p)^{1 - y/n}I_{A_2}(y),
\end{equation}
for $y = 0, 1, \cdots, n$, where $n$ is the number of trials, $p \in [0,1]$ is the constant probability of success, $\rho \in [0,1]$ is the correlation coefficient, $A_1= \{0, 1, \cdots, n\}$, and $A_2 = \{0, n\}$. As mentioned by \cite{Luce:1995}, the CB distribution is capable of more than the original binomial distribution (\ref{binomial}) in how flexible the distribution is depending on whether the Bernoulli variables have a positive or negative correlation. The logarithm of the likelihood function for the $k$ observations $\{y_1, \cdots, y_k\}$ from the CB distribution in (\ref{cb:01}) is given by
\begin{align*}
  &P( p, \rho \mid n, y_1, \cdots, y_k) \\
  & = \sum_{i = 1}^{k}\log\left\{{{n}\choose{y_i}} p^{y_i} (1 - p)^{n-y_i}(1 - \rho)I_{A_1}(y_i) + p^{y_i/n} (1-p)^{1 - y_i/n}I_{A_2}(y_i)\right\}.
\end{align*}
Due to the complexity of the likelihood function above, it may be impossible to derive explicit expressions of the MLEs for the parameters $p$ and $\rho$. Numerical approximation methods, such as the Newton–Raphson-type method, are usually required to obtain the MLEs under this scenario. However, these methods are very sensitive to starting values of the parameters. To tackle with this difficulty, \cite{Dini:Tuti:2010} considered Bayesian analysis with informative priors for the parameters of the CB distribution and developed the Metropolis-within-Gibbs algorithm for making posterior inference from the conditional posterior distributions of the unknown parameters, whereas Bayesian sensitivity analysis of prior elicitation is often necessary to evaluate bias due to misclassification of using informative priors. In this paper, to reduce the burden of prior elicitation of Bayesian analysis, we develop an expectation maximization (EM) algorithm for computing the MLEs of the CB parameters. Numerical results from simulation studies and a real-data application show that the performance of the proposed EM algorithm is quite effective in terms of the average bias and the root mean square error.% consistently reaching a global maximum.

The remainder of this paper is organized as follows. In Section \ref{section:02}, we develop an EM algorithm for computing the MLEs of the parameters of the CB distribution. In Section \ref{section:03}, we carry out numerical studies to the performance of the proposed EM-type MLE method. Finally, some concluding remarks are provided in Section \ref{section:04}, with the R programs deferred to the Appendix.

\section{The proposed EM algorithm } \label{section:02}

The EM algorithm, originally developed by \cite{Demp:Lair:1977}, is a numerical iterative algorithm to calculate the MLEs of the unknown model parameters. It consists of two steps which are (i) an expectation step (E-step) and (ii) a maximization step (M-step). It is often employed to obtain the MLEs of the parameters from a complex likelihood function of the model by iterating two above mentioned steps. To be more specific, in the E-step, we compute the expectation of the log-likelihood function under the incomplete data given the observed data, and in the M-step, we calculate the maximizer of the expected log-likelihood function. We here refer the interested readers to \cite{Demp:Lair:1977} and \cite{Litt:Rubi:1987} in detail.

First, to easily construct the EM algorithm for estimating the MLEs of the CB parameters, we introduce a latent variable $Z_i$ to indicate each component of the CB distribution $y_i, i=1, \cdots,k$ belonging to either the modified Bernoulli or the Binomial component, namely,
\begin{equation} \label{def:z}
    Z_i=
    \begin{cases}
      1, & \text{if~the~observation}~y_i~\text{is~from~modified~the~Bernoulli~distribution} \\
      0, & \text{if~the~observation}~y_i~\text{is~from~the~Binomial~distribution}
    \end{cases}.
  \end{equation}
The probability of the success of the variable $Z_i$ conditional on the $i$th component $y_i$  of the CB distribution is given by
\begin{align} \nonumber
  \tau_i &= P(Z_i = 1 \mid Y_i = y_i) = \frac{P(Y_i = y_i \mid Z_i = 1)P(Z_i = 1)}{P(Y_i = y_i)} \\
  &= \frac{p^{y_i/n} (1-p)^{1 - y_i/n}I_{A_2}(y_i)}{{{n}\choose{y}} p^y (1 - p)^{n-y}(1 - \rho)I_{A_1}(y) + p^{y/n} (1-p)^{1 - y/n}I_{A_2}(y)}. \label{taui}
\end{align}
Then, we can incorporate the above latent variable $Z=(Z_1, \cdots, Z_k)'$ to construct the logarithm of the complete-data likelihood function given by
\begin{align}\nonumber
 & \log L(\theta \mid n, y, z) \\ \nonumber
  &= \sum_{i=1}^{k} z_i \log(\rho) + \sum_{i=1}^{k}(1-z_i)\log(1-\rho) + \sum_{i=1}^{k}\left[ \frac{z_iy_i}{n} + (1-z_i)y_i\right]\log(p) \\
  &= \sum_{i=1}^{k}\left[ \frac{z_i(n-y_i)}{n} + (1-z_i)(n-y_i)\right]\log(1-\rho) + \sum_{i=1}^{k}(1-z_i)\log{n \choose y_i}, \label{logcomplete}
\end{align}
where $\theta = (p, \rho)$. For notational simplicity, let $\theta^{(t)} = (p^{(t)}, \rho^{(t)})$ be an estimate of $\theta$ at the $t$th EM sequence. In what follows, using Equation (\ref{taui}), we summarize the EM algorithm for obtaining the MLEs of the parameters for the CB distribution as follows.
\begin{itemize}
  \item Estimation Step

  Using Equation (\ref{logcomplete}), we obtain
  \begin{align} \nonumber
   Q(\theta \mid \theta^{(t)}) &= \mathrm{E}\left[\log L(\theta \mid n, y, z) \mid \theta^{(t)} \right]\\ \nonumber
    &= \sum_{i=1}^{k}\mathrm{E}(Z_i \mid \theta^{(t)})\log(\rho) + \sum_{i=1}^{k}\left[1-\mathrm{E}(Z_i \mid \theta^{(t)})\right]\log(1 - \rho)\\ \nonumber
    & + \sum_{i=1}^{k}\left[ \frac{\mathrm{E}(Z_i \mid \theta^{(t)})y_i}{n} + (1-\mathrm{E}(Z_i \mid \theta^{(t)}))y_i\right]\log(p)\\ \nonumber
    & + \sum_{i=1}^{k}\left[ \frac{\mathrm{E}(Z_i \mid \theta^{(t)})(n-y_i)}{n} + (1-\mathrm{E}(Z_i \mid \theta^{(t)}))(n-y_i)\right]\log(1-\rho)\\ \label{taui:est1}
    & +  \sum_{i=1}^{k}(1-\mathrm{E}(Z_i \mid \theta^{(t)}))\log{n \choose y_i}.
  \end{align}
  Based on the latent variable $Z_i$ defined in (\ref{def:z}) and Equation (\ref{taui}), we obtain that
\begin{align}\nonumber
 \hat{\tau}_i &= \mathrm{E}(Z_i \mid \theta^{(t)}) \\ \label{taui:est}
  & = \frac{p^{(t)y_i/n} (1-p^{(t)})^{1 - y_i/n}I_{A_2}(y_i)}{{{n}\choose{y}} p^{(t)y} (1 - p^{(t)})^{n-y}(1 - \rho^{(t)})I_{A_1}(y) + p^{(t)y/n} (1-p^{(t)})^{1 - y/n}I_{A_2}(y)}.
\end{align}
It is immediate upon substituting Equation (\ref{taui:est}) into (\ref{taui:est1}) that we have
  \begin{align*}
  & Q(\theta \mid \theta^{(t)}) = \mathrm{E}\left[\log L(\theta \mid n, y, z) \mid \theta^{(t)} \right]\\
    &= \sum_{i=1}^{k}\theta^{(t)}\log(\rho) + \sum_{i=1}^{k}\left[1-\theta^{(t)}\right]\log(1 - \rho) + \sum_{i=1}^{k}\left[ \frac{\theta^{(t)}y_i}{n} + (1-\theta^{(t)})y_i\right]\log(p)\\
    & + \sum_{i=1}^{k}\left[ \frac{\theta^{(t)}(n-y_i)}{n} + (1-\theta^{(t)})(n-y_i)\right]\log(1-\rho) +  \sum_{i=1}^{k}(1-\theta^{(t)})\log{n \choose y_i}.
  \end{align*}

  \item Maximization Step

  We take differentiate $Q(\theta \mid \theta^{(t)})$ with respect to the parameters $p$ and $\rho$ in Equation (9) and obtain that
  \begin{equation}
    \begin{cases}
      \frac{\partial Q(\theta \mid \theta^{(t)})}{\partial p}    = \frac{\sum_{i=1}^{k}\left[ \tau_i y_i/n + (1 - \tau_i)y_i\right]}{p} -  \frac{\sum_{i=1}^{k}\left[ (n-y_i)\tau_i/n + (n - y_i)(1-\tau_i)\right]}{1-p}\\
      \frac{\partial Q(\theta \mid \theta^{(t)})}{\partial \rho}  = \frac{\sum_{i=1}^{k}\tau_i}{\rho} - \frac{\sum_{i=1}^{k}(1-\tau_i)}{1 - \rho}
    \end{cases}.
  \end{equation}
  By setting the two equations above to be zero to solve for $\rho$ and $p$, we obtain the $(t+1)$th EM sequence, denoted by  $\theta^{(t+1)} = (p^{(t+1)}, \rho^{(t+1)})$, such that
  $$
  \hat{\rho} = \frac{\sum_{i=1}^{k}\hat{\tau}_i}{k} \ \ \mathrm{and} \ \ \hat{p} = \frac{\sum_{i=1}^{k}\left[ \hat{\tau}_i y_i/n + (1 - \hat{\tau}_i)y_i\right]}{\sum_{i=1}^{k}\left[ n(1 - \hat{\tau}_i) + \hat{\tau}_i\right]},
  $$
  which have simple analytical expressions. Thus, these expressions allow easy calculation of the MLEs for the parameters of the CB distribution based on the EM algorithm above. To make our algorithm accessible to practitioners, we provide the R code in Appendix and utilize a real-data example to demonstrate how to implement the proposed EM algorithm.
\end{itemize}

\section{Numerical results} \label{section:03}

In this section, we first carry out simulation studies to investigate the performance of the EM algorithm for estimating the MLEs of the BC parameters under different scenarios in Section \ref{section:03:01} and then utilize a real-data example for illustrative purposes in Section \ref{section:03:02}.

\subsection{Simulation studies} \label{section:03:01}

\begin{figure}[thb]
\centering
\includegraphics[width=70mm] {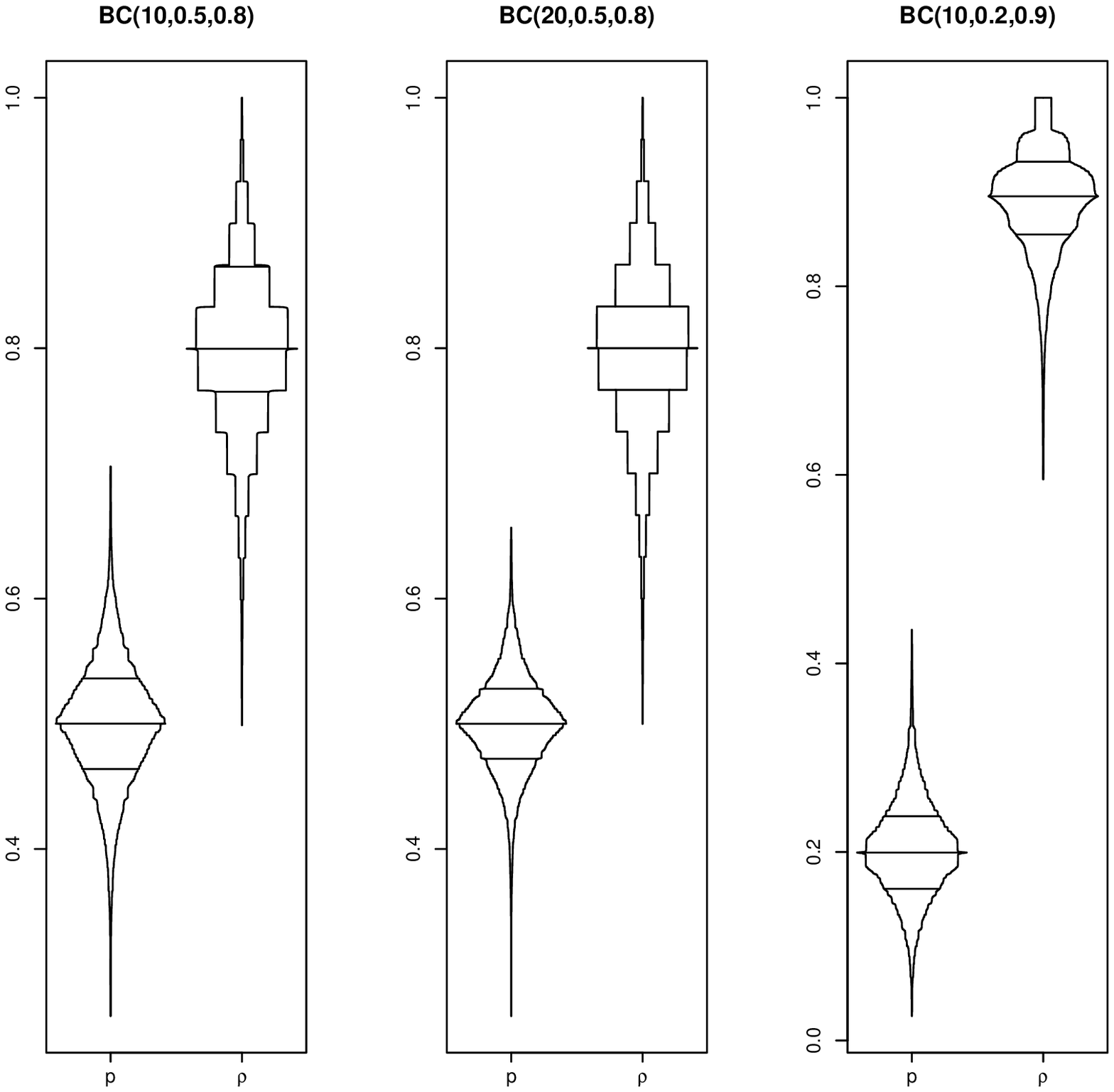}  %%% NEW GRAPH
\includegraphics[width=70mm] {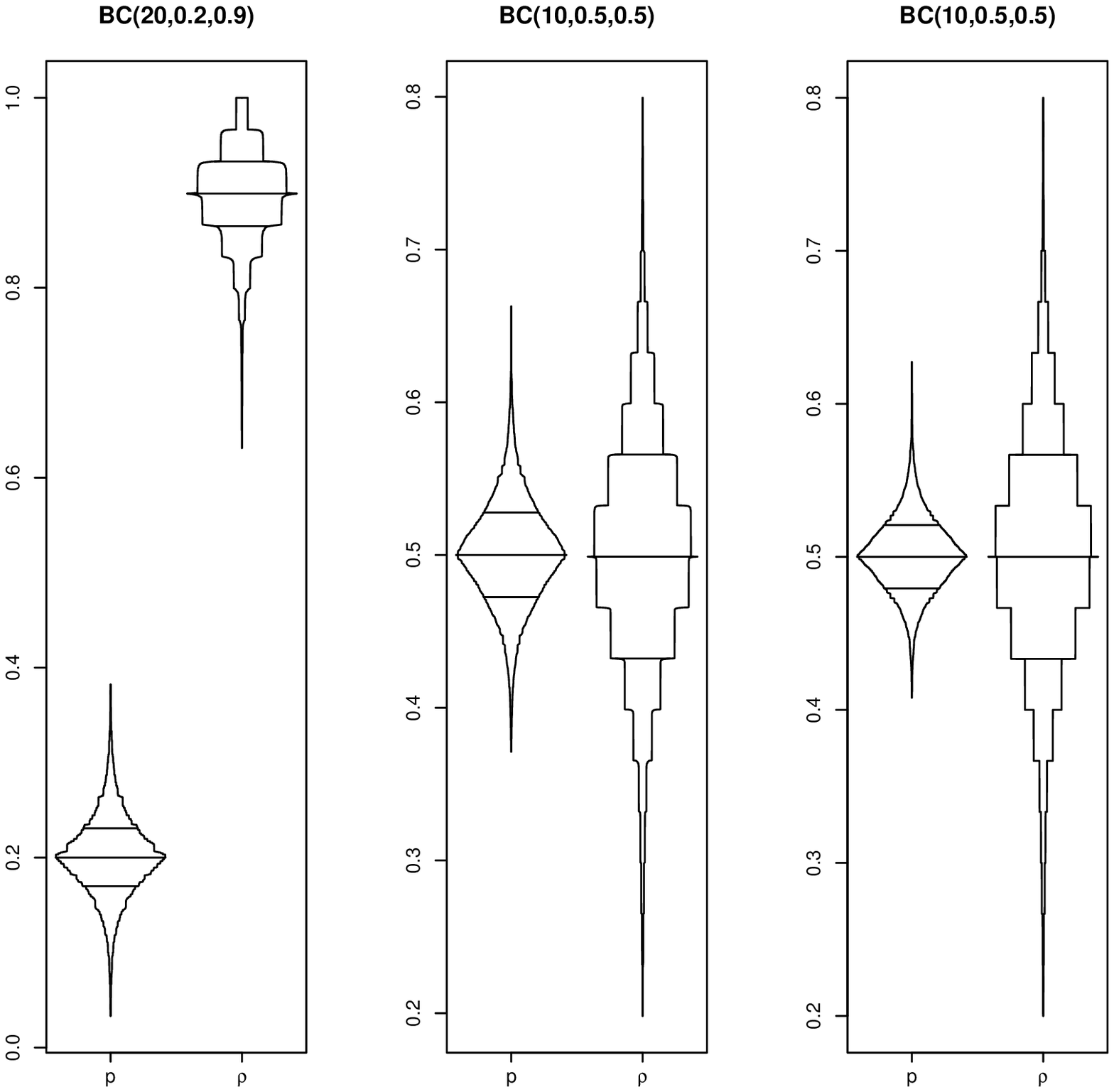}
\vspace{-2.0\baselineskip}
\caption{Box-percentile plots of the estimates of the two parameters $p$ and $\rho$ using the proposed EM algorithm based on $1,000$ replications.} \label{figure1}
\end{figure}

We generate random samples of size $k = 30$ from the BC distribution in (\ref{cb:01}) with $(n, p, \rho)= \{ (10,0.5,0.8), (20,0.5,0.8),(10,0.2,0.9),(20,0.2,0.9),(10,0.5,0.5),(20,0.5,0.5) \}$. For each simulation setting, we replicate $N = 1,000$ times and compute the average bias and root mean square error (RMSE) of each estimator given by
$$
\mathrm{Bias} = \frac{\sum_{i=1}^{N}(\hat{\theta}_i - \theta)}{N} \ \ \mathrm{and} \ \ \mathrm{RMSE} = \sqrt{\frac{\sum_{i=1}^{N}(\hat{\theta}_i - \theta)^2}{N}},
$$
where $\hat{\theta}_i$ stands the MLE for the true parameter $\theta_i$ based on the EM algorithm in the $i$th replication. In addition, we calculate the 95\% confidence coverage of each parameter based on M replications. We report numerical results in Table \ref{table1} and depict the box-percentile plots of the EM-type MLE estimates under each scenario in Figure \ref{figure1}, which could allow researchers to easily visualize the shape of the estimate of each parameter based on $N$ replications. Several findings can be summarized as follows: (i) The proposed EM algorithm always provides reliable estimates for the two parameters in terms of the bias and RMSE. (ii) The bias and RMSE decrease as the number of trials $n$ increases. (iii) The estimate of $\rho$ shows a greater variability than the estimate of $p$, shown in Figure \ref{figure1}. Finally, it deserves mentioning that we have conducted extensive simulation studies under other simulation scenarios and achieved the same conclusions, but due to the space limitations, we omitted these results here. In summary, we highly recommend the use of the proposed EM algorithm for obtaining the MLEs of the parameters of the BC distribution.

\begin{center}
\begin{table}
\caption{The average bias and RMSE of the estimates for the parameters $p$ and $\rho$ under different simulation scenarios based on $1,000$ replications.} \label{table1}
\begin{tabular}{@{}ccccccc@{}}
\hline
&& & & \\
Model &Parameter &
\multicolumn{1}{c}{Bias} &
RMSE&
\multicolumn{1}{c}{95\% frequentist coverage}  \\
\hline
{BC(10, 0.5, 0.8)} & $p$       & 0.0008097407    & 0.05771765 & [0.3849653, 0.6093339]  \\
                 & $\rho$    & $-$0.0016802761 & 0.30470560 & [0.6325764, 0.9331690]    \\[6pt]
{BC(20, 0.5, 0.8)} & $p$       & 0.0005851561	& 0.04473148	&[0.4137872, 0.5889664]  \\
                 & $\rho$    & $-$0.0014346316&	0.30273720&	[0.6333326, 0.9333331]    \\[6pt]
{BC(10, 0.2, 0.9)} & $p$       & 0.000899965	&0.05855643&	[0.08966766, 0.3306402]  \\
                 & $\rho$    & $-$0.004418155	&0.70154750&	[0.75164919 1.0000000]    \\[6pt]
{BC(20, 0.2, 0.9)} & $p$       & 0.0004018936	&0.04758628&	[0.1100396,   0.3]  \\
                 & $\rho$    & $-$0.0020235147	&0.70121450&	[0.7659021,   1.0]    \\[6pt]
{BC(10, 0.5, 0.5)} & $p$       & 0.0008298881	&0.04078532&	[0.4217728, 0.5803682]  \\
                 & $\rho$    & 0.0016500196&	0.04078532&	[0.3315481, 0.6659845]    \\[6pt]
{BC(20, 0.5, 0.5)} & $p$       & 0.0007427234	&0.02914349&	[0.4475884, 0.5587433]  \\
                 & $\rho$    & 0.002131605	&0.02914349&	[0.3333316, 0.6666659]    \\[6pt]
\hline
\end{tabular}
\end{table}
\end{center}

\subsection{A real-data application} \label{section:03:02}

In this section, we consider a real data application, shown in Table \ref{sphericcase}, about a commercial soybean selection study analyzed in \cite{Dini:Tuti:2010} to illustrate the effectiveness of the proposed EM algorithm. We refer the readers to \cite{Dini:Tuti:2010} for more details in the description of this IAC23 soybean study.

\begin{center}
\begin{table}[htbp]
\caption{The number of IAC23 soybean plants selected in each plot after 15 days.} \label{sphericcase}
\begin{tabular}{cccccccccc}
 \hline
4 & 4 & 6 & 2 & 3 & 3 & 3 & 5 & 5 & 6\\
6 & 3 & 3 & 4 & 1 & 1 & 5 & 4 & 4 & 2\\ \hline
\end{tabular}
\end{table}
\end{center}

Specifically, we observe that \cite{Dini:Tuti:2010} considered Bayesian methods for estimating the parameters of the CB distribution by specifying different priors for $p$ and $\rho$. For illustrative purposes, we here consider the uniform priors for the two parameters to represent prior ignorance to drive objective Bayesian inference. The resulting posterior means for the two parameters are given by $\hat{p}=0.5826$ and $\hat{\rho}=0.1296$, respectively. The corresponding likelihood and log-likelihood values are 1.344801e-16 and $-36.54512$, respectively. We reanalyze the above real data with the proposed EM algorithm in Section \ref{section:02}. For this data set, we obtained that $\hat{p}=0.5869412$ and $\hat{\rho}=0.0863572$. The corresponding likelihood and log-likelihood values are 1.491578e-16 and $-36.44153$, respectively. By comparing the two results above, we may conclude that the estimates based on the proposed EM algorithm have a greater likelihood value, indicating that it is more effective than Bayesian estimations by \cite{Dini:Tuti:2010}.

\section{Concluding remarks} \label{section:04}

In this paper, we proposed an EM algorithm for obtaining the MLEs of the two parameters of the correlated binomial model, an alternative to the binomial distribution for the analysis of the data in the presence of correlations among events. Numerical studies showed that the proposed method is more effective by consistently reaching a global maximum. More importantly, the proposed EM algorithm is not only easily implemented by practitioners with the R programs in the Appendix, but also can be easily taught to senior undergraduate or graduate  students with an emphasis on the interested applications of the EM algorithm for computing the MLEs of the parameters in discrete mixture models.

\newpage
\section*{Appendix. R programs for the proposed EM algorithm}
\begin{quote}
\begin{small}
%%\noindent\rule{\textwidth-\leftmargin-\rightmargin}{0.5pt}
\hrulefill
\begin{verbatim}
#CB.EM function
CB.EM = function(y, n, start = c(0.5, 0.5), maxits = 1000 L, eps = 1.0E-15){
  k = length(y)
  converged1 = logical(k)
  converged2 = logical(k)
  z = rep(0, k)
  index = c(which(y == n), which(y == 0))
  theta = start
  p = start[1]
  rho = start[2]
  iter = 1
  for (i in index) {
    p1 = rho * p ^ (y[i] / n) * (1 - p) ^ ((n - y[i]) / n)
    p2 = (1 - rho) * dbinom(y[i], n, p, log = FALSE)
    z[i] = p1 / (p1 + p2)
  }
  p = sum(z * y / n + (1 - z) * y) / sum(z + (1 - z) * n)
  rho = mean(z)
  theta = c(p, rho)
  while ((iter < maxits) && (!converged1) && (!converged2)) {
    for (i in index) {
      p1 = rho * p ^ (y[i] / n) * (1 - p) ^ ((n - y[i]) / n)
      p2 = (1 - rho) * dbinom(y[i], n, p)
      z[i] = p1 / (p1 + p2)
    }
    p = sum(z * y / n + (1 - z) * y) / sum(z + (1 - z) * n)
    rho = mean(z)
    newtheta = c(p, rho)
    converged1 = abs(newtheta[1] - theta[1]) < eps
    converged2 = abs(newtheta[2] - theta[2]) < eps
    iter = iter + 1 L
    theta = newtheta
  }
  list(para = theta, iter = iter, conv = c(converged1, converged2))
}
\end{verbatim}
%-------------
\hrulefill
%-------------
\begin{color}{red}
\begin{verbatim}
#Data
> y = c(4, 4, 6, 2, 3, 3, 3, 5, 5, 6, 6, 3, 3, 4, 1, 1, 5, 4, 4, 2)
\end{verbatim}
\end{color}
%-------------
\hrulefill
%-------------
\begin{color}{red}
\begin{verbatim}
# Use of CB.EM() function
> CB.EM(y, n=6, start = c(0.5, 0.1), maxits = 1000 L, eps = 1.0E-15)
$para
  [1] 0.5869412 0.0863572
$iter
  [1] 55
$conv
  [1] TRUE TRUE
\end{verbatim}
\end{color}
\hrulefill
\end{small}
\end{quote}

%% or include bibliography directly:
\bibliographystyle{annals}

\end{document}